\documentclass[12pt]{article}
\usepackage{amssymb}
\usepackage{epsfig}
\usepackage{epsfig,bbm}

\def\hybrid{\topmargin -20pt    \oddsidemargin 0pt
        \headheight 0pt \headsep 0pt
        \textwidth 6.25in       
        \textheight 9.5in       
        \marginparwidth .875in
        \parskip 5pt plus 1pt   \jot = 1.5ex}

\hybrid

\def\baselinestretch{1.2}

\catcode`\@=11

\def\marginnote#1{}
\newcount\hour
\newcount\minute
\newtoks\amorpm
\hour=\time\divide\hour by60 \minute=\time{\multiply\hour by60
\global\advance\minute by-\hour}
\edef\standardtime{{\ifnum\hour<12 \global\amorpm={am}%
        \else\global\amorpm={pm}\advance\hour by-12 \fi
        \ifnum\hour=0 \hour=12 \fi
        \number\hour:\ifnum\minute<10 0\fi\number\minute\the\amorpm}}
\edef\militarytime{\number\hour:\ifnum\minute<10
0\fi\number\minute}

\def\draftlabel#1{{\@bsphack\if@filesw {\let\thepage\relax
   \xdef\@gtempa{\write\@auxout{\string
      \newlabel{#1}{{\@currentlabel}{\thepage}}}}}\@gtempa
   \if@nobreak \ifvmode\nobreak\fi\fi\fi\@esphack}
        \gdef\@eqnlabel{#1}}
\def\@eqnlabel{}
\def\@vacuum{}
\def\draftmarginnote#1{\marginpar{\raggedright\scriptsize\tt#1}}

\def\draft{\oddsidemargin -.5truein
        \def\@oddfoot{\sl preliminary draft \hfil
        \rm\thepage\hfil\sl\today\quad\militarytime}
        \let\@evenfoot\@oddfoot \overfullrule 3pt
        \let\label=\draftlabel
        \let\marginnote=\draftmarginnote
   \def\@eqnnum{(\theequation)\rlap{\kern\marginparsep\tt\@eqnlabel}%
\global\let\@eqnlabel\@vacuum}  }

\def\preprint{\twocolumn\sloppy\flushbottom\parindent 2em
        \leftmargini 2em\leftmarginv .5em\leftmarginvi .5em
        \oddsidemargin -.5in    \evensidemargin -.5in
        \columnsep .4in \footheight 0pt
        \textwidth 10.in        \topmargin  -.4in
        \headheight 12pt \topskip .4in
        \textheight 6.9in \footskip 0pt
        \def\@oddhead{\thepage\hfil\addtocounter{page}{1}\thepage}
        \let\@evenhead\@oddhead \def\@oddfoot{} \def\@evenfoot{} }

\def\numberbysection{\@addtoreset{equation}{section}
        \def\theequation{\thesection.\arabic{equation}}}

\def\underline#1{\relax\ifmmode\@@underline#1\else
        $\@@underline{\hbox{#1}}$\relax\fi}

\def\titlepage{\@restonecolfalse\if@twocolumn\@restonecoltrue\onecolumn
     \else \newpage \fi \thispagestyle{empty}\c@page\z@
        \def\thefootnote{\fnsymbol{footnote}} }

\def\endtitlepage{\if@restonecol\twocolumn \else \newpage \fi
        \def\thefootnote{\arabic{footnote}}
        \setcounter{footnote}{0}}  

\catcode`@=12 \relax

\def\figcap{\section*{Figure Captions\markboth
        {FIGURECAPTIONS}{FIGURECAPTIONS}}\list
        {Figure \arabic{enumi}:\hfill}{\settowidth\labelwidth{Figure
999:}
        \leftmargin\labelwidth
        \advance\leftmargin\labelsep\usecounter{enumi}}}
 \relax
\def\tablecap{\section*{Table Captions\markboth
        {TABLECAPTIONS}{TABLECAPTIONS}}\list
        {Table \arabic{enumi}:\hfill}{\settowidth\labelwidth{Table
999:}
        \leftmargin\labelwidth
        \advance\leftmargin\labelsep\usecounter{enumi}}}
 \relax
\def\reflist{\section*{References\markboth
        {REFLIST}{REFLIST}}\list
        {[\arabic{enumi}]\hfill}{\settowidth\labelwidth{[999]}
        \leftmargin\labelwidth
        \advance\leftmargin\labelsep\usecounter{enumi}}}
 \relax
\makeatletter
\newcounter{pubctr}
\def\publist{\@ifnextchar[{\@publist}{\@@publist}}
\def\@publist[#1]{\list
        {[\arabic{pubctr}]\hfill}{\settowidth\labelwidth{[999]}
        \leftmargin\labelwidth
        \advance\leftmargin\labelsep
        \@nmbrlisttrue\def\@listctr{pubctr}
        \setcounter{pubctr}{#1}\addtocounter{pubctr}{-1}}}
\def\@@publist{\list
        {[\arabic{pubctr}]\hfill}{\settowidth\labelwidth{[999]}
        \leftmargin\labelwidth
        \advance\leftmargin\labelsep
        \@nmbrlisttrue\def\@listctr{pubctr}}}
 \relax
\makeatother
\newskip\humongous \humongous=0pt plus 1000pt minus 1000pt

\newif\ifdtup

\relax

\def\be{\begin{equation}}
\def\ee{\end{equation}}
\def\ba{\begin{eqnarray}}
\def\ea{\end{eqnarray}}

\def\a{\alpha}

\def\b{\beta}

\def\g{\gamma}

\def\d{\delta}

\def\e{\epsilon}

\def\l{\lambda}

\def\no{\noindent}

\def\IR{\relax{\rm I\kern-.18em R}}

\def\IR{\relax{\rm I\kern-.18em R}}
\def\inv{^{\raise.15ex\hbox{${\scriptscriptstyle -}$}\kern-.05em 1}}

\def\R{{\mathbb{R}}}

\begin{document}

\renewcommand{\theequation}{\thesection.\arabic{equation}}

\newcommand{\beq}{\begin{equation}}
\newcommand{\eeq}[1]{\label{#1}\end{equation}}
\newcommand{\ber}{\begin{eqnarray}}
\newcommand{\eer}[1]{\label{#1}\end{eqnarray}}
\newcommand{\eqn}[1]{(\ref{#1})}
\newcommand{\pa}{\partial}

\begin{titlepage}
\begin{center}

\vskip -.1 cm
\hfill March 2009\\

\vskip .4in

{\large \bf Critical behavior of collapsing surfaces}

\vskip 0.6in

{\bf Kasper Olsen}
\vskip 0.1in
{\em Department of Physics, Technical University of Denmark\\
DK-2800 Lyngby, Denmark

\vskip 0.1in
\footnotesize{\tt kasper.olsen(at)fysik.dtu.dk}}\\

\vskip 0.2in

and

\vskip 0.2in

{\bf Christos Sourdis}
\vskip 0.1in
{\em Department of Physics, University of Patras \\
GR-26500 Patras, Greece\\

\vskip 0.1in
\footnotesize{\tt sourdis(at)pythagoras.physics.upatras.gr}}\\

\end{center}

\vskip 0.8in

\centerline{\bf Abstract}
\no
We consider the mean curvature evolution of rotationally symmetric surfaces. Using numerical
methods, we detect critical behavior at the threshold of singularity formation resembling the one of gravitational collapse. In particular, the mean curvature simulation of a one-parameter family of initial data reveals the existence of a critical initial surface that develops a degenerate neckpinch. The limiting flow of the Type II singularity is accurately modeled by the rotationally symmetric translating soliton.

\vfill
\end{titlepage}
\eject

\def\baselinestretch{1.2}
\baselineskip 16 pt \noindent

\section{Introduction}

Geometric evolution equations  are diffusive (heat type) equations that describe the deformation of metrics on Riemannian manifolds driven by their curvature in various forms.  These deformations are naturally divided into two classes that correspond to intrinsic and extrinsic curvature flows. The former class refers to deformations driven by the intrinsic Ricci curvature tensor on a manifold, whereas the latter refers to deformations of submanifolds embedded in higher-dimensional spaces evolving by their extrinsic curvature. The field of geometric flows is a subject of interest in both  physics and mathematics with important recent advances allowing for further striking developments.

The understanding - and in many cases the mathematical classification - of  the formation of possible singularities along different flows has played an important role in the progress of the field of geometric flows. The authors of \cite{G-I 2003}, by numerically simulating the Ricci
flow of a specific one-parameter family of rotationally symmetric geometries on $S^3$ with a varying amount of $S^2$ neckpinching, found critical behavior at the threshold of singularity
formation. According to their results, the Ricci flow for a critical initial geometry, i.e., for the transition point between initial geometries on $S^3$ which under the volume normalized flow converge to a sphere and other initial geometries that develop a singular neckpinch, evolves into a degenerate neckpinch. In addition, in \cite{G-I 2007} it was shown that the limiting flow of the critical geometry is modeled by the Bryant Ricci soliton.

In this work we initiate the search for critical behavior of singularity formation of extrinsic curvature flows. With the purpose of revealing a critical behavior analogues to the one close to the threshold of black hole formation \cite{Choptuik}, \cite{Gundlach}, we perform a numerical study of the singularity structure of closed rotationally symmetric surfaces evolving under their mean curvature.

In order to investigate whether critical behavior is present in extrinsic curvature flows, we follow a similar numerical approach to the one of \cite{G-I 2003, G-I 2007} adjusted to the mean curvature flow.  The plan of the paper is as follows: section 2 starts by briefly reviewing the mean curvature evolution of surfaces of revolution and commenting on the classification of their singularity pattern. Then, a one-parameter family of surfaces with a dumbbell-like shape is introduced to be used in the current study. Emphasis is placed upon the two-dimensional mean curvature translating soliton and its role in Type II singularity formation. In section 3, the numerical results are presented. These results provide strong evidence that critical behavior is indeed detected in the simulation of the mean curvature evolution of the one-parameter family of initial data of our choice. The limiting flow of a critical initial surface is shown to be accurately modeled by the appropriate mean curvature steady soliton. Furthermore, in the same section, it is shown that the  critical surface develops a degenerate neckpinch at the final stage of its evolution. Where appropriate, frequent references are made to existing mathematical results that formed the original basis of the expectations for mean curvature critical behavior. Finally, section 4 presents both the conclusions and possible future research directions related to critical phenomena of gravitational collapse.

\section{Mean curvature flow}

A smooth $n$-dimensional hypersurface embedded in Euclidean space, $\vec{{}X}: M^n \to \R^{n+1}$,
evolves by its mean curvature if the normal velocity of any point $ p\in M^n$, at the time of evolution $t$, coincides with the mean curvature $H(p,t)$ at that point,
\be
\frac{\partial {{\vec{{}X}}}(p,t)}{\partial t} = -H(p,t)\,{\hat n}(p,t)\,,
\label{mcf} 
\ee
where ${\hat n}$ is the outer unit normal vector at $\vec{{}X}(p,t)$.\footnote{The signs are chosen
such that $\vec{{{}H}} = -H{\hat n}$ is the mean curvature vector with the mean curvature of a
 convex hypersurface being positive.} In a local coordinate system  the metric and the second fundamental form induced on $M_t = \vec{{}X_t}(M^n)$ are given respectively by
\be
g_{\a\b}(p,t) = \partial_\a {{\vec{{}X}}}(p,t) {\cdot}\, \partial_\b {{\vec{{}X}}}(p,t)\,,
\ee
and
\be
h_{\a\b}(p,t) = -{\hat n}(p,t) {\cdot}\, \partial_{\a\b}^2 {{\vec{{}X}}}(p,t)\,,
\ee
with $1\le \a,\b \le n$. The eigenvalues of the second fundamental form $h_{\a\b}$ with respect to
the induced metric $g_{\a\b}$, usually denoted as $\kappa_\a$, are the principal curvatures of the hypersurface $M^n$. The mean curvature is given by the trace of the second fundamental form $H = g^{\a\b}h_{\a\b} = \kappa_1 + \kappa_2+ \ldots \kappa_n$, and is related to the scalar curvature $R$ according to
\be
R(p,t) =  H^2(p,t) -|A(p,t)|^2\,,
\ee
where $|A|^2 = g^{\a\b}g^{\g\d}h_{\a\g}h_{\b\d} = \kappa_1^2 + \kappa_2^2 + \ldots + \kappa_n^2$.

In the following sections, and after first specializing on the particular surfaces of interest, the differential equations that govern
the evolution of the geometric quantities just introduced are used. These are the time evolution equations of $g_{\a\b}$, $h_{\a\b}$ and $H$ (that can be found in the introductory chapters of any textbook on mean curvature flow, e.g. \cite{mcfBs}),
\ba
\frac{\pa\, g_{\a\b}}{\pa t} &=& -2 H h_{\a\b}\,,\label{mcfg}\\
\frac{\pa h_{\a\b}}{\pa t} &=& \Delta h_{\a\b} -2 H h_{\a\g}{h^\g}_\b + |A|^2 h_{\a\b}\,,\label{mcfh}\\
\frac{\pa H}{\pa t}\,{}\, &=& \Delta H + |A|^2 H\label{mcfH}\,,
\ea
where the dependence of all quantities involved on $p$ and $t$ has been suppressed.

Under the mean curvature flow (\ref{mcf}), a compact and closed  hypersurface $M_0$ of positive mean curvature embedded in Euclidean space gets extinct after a maximal finite time $T_*$ subject to the bounds
\cite{giga-yama, evans-spruck, huisken-sine99}
\be
2\,\frac{{{\cal V}(M_0)}^2}{{{\cal A}(M_0)}^2} \le\, T_*\, \le \, \frac{{\mbox{diam}}(M_0)^2}{8n}\,,
\label{bounds}
\ee
given here in terms of the volume ${\cal V}(M_0)$ enclosed by the initial $n$-dimensional hypersurface, its initial area ${\cal A}(M_0)$ and diameter. In the course of its evolution the hypersurface may possibly become singular at an earlier time $T < T_*$.  In this case, the solution $M_t$ of the mean curvature flow (\ref{mcf}) for the initial hypersurface $M_0$ exhibits a singularity before its collapse to one  point (or possibly more), that is classified as \textbf{Type I} or \textbf{II} according to the rate at which the maximal curvature $A_{max}(t) = \displaystyle \max_{M_t} A(t)$
blows up as $t \to T$. If  the maximal  curvature can be shown to obey
\be
A_{max}(t) \leq \frac{C} {\sqrt {T - t}}\,,
\label{Arate}
\ee
for some constant $C < \infty$, then $M_t$ exhibits a Type I singularity. Otherwise, it is said that a Type II singularity is forming as $t \to T$.
Compared to the curve shortening flow, note that singularities of the second type, or ``slowly forming" singularities, do not occur during the evolution of compact curves embedded on the Euclidean plane. The final stage of such curves moving under  their mean curvature is to become convex - if they were not initially - and shrink to a point as circles \cite{grayson}. Similarly, compact convex hypersurfaces asymptotically converge to the self-similar round sphere solution and collapse to a point of $\R^{n+1}$ with the formation of a Type I singularity \cite{huisken:1984}.

The prototype example of a flow that is commonly invoked to argue in support of the possibility that an embedded hypersurface may become singular before it shrinks to a point is provided  by an appropriate dumbbell in $\R^3$, see for instance \cite{Angenent:Tori}.  Consider a dumbbell consisting of two large approximately round spheres connected by a sufficiently thin cylindrical neck as initial configuration. Due to its large curvature, the neck pitches and a singularity is formed before the two spheres have the chance to shrink considerably. However, if the neck is not ``very thin",  the dumbbell surface becomes convex in the course of its flow and, in accordance to the standard theorem of \cite{huisken:1984}, collapses to a single point. An interesting intermediate scenario is also possible. There exist dumbbell-shaped surfaces with just the right neck that shrink to a point without ever turning convex nor loosing their neck under their mean curvature evolution.  This specific class of mean curvature flow solutions that exhibit precisely this degenerate neckpinching behavior is the primary interest of the current work. The maximal curvature for these solutions does not satisfy (\ref{Arate}) for any finite constant $C$, thus the forming singularity at their collapse is of Type II. The existence of such mean curvature flow solutions was originally suggested by R. Hamilton \cite{AAG}. He indicated that solutions with a maximal curvature blowing up faster than the rate (\ref{Arate}) can be produced by studying rotationally symmetric surfaces that exhibit non-generic neckpinching. These solutions were investigated using level-set methods initially in \cite{evans-spruck} and later in \cite{AAG} where a proof of the Type II character of the forming singularity was presented. A detailed study of the possible blow-ups of these solutions can be found in \cite{angenent-velaz} along with a piecewise construction of the corresponding solutions.

It should be noted here that the above discussion refers to left-right symmetric dumbbell-shaped surfaces, in accordance with the choice of surfaces to be investigated in the sequel. Reflection non-symmetric surfaces with an intermediate, or ``critical" in the nomenclature of the present work, initial data may also exhibit similar singular behavior \cite{AAG, angenent-velaz}. This discussion can be easily generalized
to hypersurfaces of dimension $n\ge 3$ and/or more than one number of initial necks \cite{AAG, angenent-velaz}. As a final point, a reference is made to \cite{angenent:1991} and \cite{altschuler:1991} for the curve shortening evolution and Type II singularity formation of planar or space not embedded curves.

\subsection{Surfaces of revolution}

In this work rotationally symmetric compact surfaces of a dumbbell shape are employed as initial configurations to study the formation of Type II mean curvature flow  singularities. The expression that a generic dumbbell-shaped surface exhibits a pinch at some time $T \le T_*$ (see equation (\ref{bounds})) is used if,  as $t \to T$, its neck  pinches off
leaving behind two ``tear" shaped components on either side of the neck. This work will not be concerned  with the subsequent evolution of these two separate components. For a rigorous presentation of the generalized solution beyond the first singular time $T$ using level-set flow methods,  see e.g. \cite{evans-spruck, AAG, Lset, Ilmanen} and for an update on surgery techniques see \cite{huiskens-sine2006}. Next, the essentials for describing the mean curvature evolution of a compact surface of revolution are introduced and an adequate as well as simple set-up is provided to accomplish the task of numerically revealing critical behavior in mean curvature flow.

A surface of revolution is produced by revolving  a planar profile curve (generator) about a line in $\R^3$ (the axis of revolution). For convenience, the profile curve is chosen to lie in the $xy$-plane of $\R^3$ and the axis of revolution to coincide with the $x$-axis. Using the principal patch of the standard parametrization of a compact surface of revolution generated by a curve $\left(x(u),y(u)\right)$,
\be
X = x(u) ~, ~~~~~
Y = y(u) ~ {\rm cos} \phi ~, ~~~~~
Z = y(u) ~ {\rm sin} \phi  ~,
\label{spuf}
\ee
with $\phi \in (0,2\pi)$ and $u$ ranging in some finite interval, the induced area element has components
\be
g_{uu} = x^\prime(u)^2 + y^\prime(u)^2 , ~~~~~
g_{\phi \phi} = y^2(u) , ~~~~~ g_{u \phi} = 0 ~,
\ee
while the unit surface normal is
\be
\hat{n} = {1 \over \sqrt{x^\prime(u)^2 + y^\prime(u)^2}}
\left(y^{\prime} (u), - x^\prime(u)\,{\rm cos} \phi , -x^\prime(u)
\,{\rm sin} \phi     \right)\,.
\ee
The mean curvature $H(u)$ of a surface of revolution is given by the sum of the principal curvature of its meridians $\kappa_u$ and the principal curvature $\kappa_\phi$ of its parallels
\ba
\kappa_u = { x^\prime(u)\,y^{\prime\prime}(u) - y^\prime(u)\,x^{\prime\prime}(u) \over \left(x^\prime(u)^2 + y^\prime(u)^2\right)^{3/2}}~,\\
\kappa_\phi = -{ x^\prime(u) \over |y(u)|\left(x^\prime(u)^2 + y^\prime(u)^2\right)^{1/2}}~.
\ea
The time evolution of a rotationally symmetric surface embedded in $\R^3$ under its mean curvature
\be
\frac{\partial {{\vec{{}X}}}(u,t)}{\partial t} = -H(u,t)\,{\hat n}(u,t)
\label{mcfS}\,, 
\ee
can effectively be encoded into the time evolution of the revolving planar
curve that generates it. This way, the mean curvature flow in $\R^3$ is reduced to a planar deformation of
a certain kind for the profile curve described as a graph form in the following subsection.

Under mean curvature flow,  an initially rotationally symmetric surface $M_0$ remains rotationally symmetric in the course of its regular evolution $M_t$. The evolution of the induced metric is dictated by (\ref{mcfg}) in terms of the mean curvature and the components of the second fundamental form. More precisely, if the non-zero components of the two-dimensional metric are denoted as
\ba
g_{uu}(u,t) &\equiv& {S^\prime}^2(u,t) + {R^\prime}^2(u,t)\,, \\
g_{\phi \phi}(u,t) &\equiv&  R^2(u,t)~,
\ea
then it can easily be verified that the flow equations (\ref{mcfg}) for the metric of a surface of revolution  can be described by the planar evolution
\ba
\frac{\pa R}{\pa t}&=& \frac{{S^\prime}}{\left({S^\prime}^2 + {R^\prime}^2\right)^2}\,\left( {S^\prime}{R^{\prime\prime}}-{R^\prime}{S^{\prime\prime}}\right) - \frac{{S^\prime}^2}{R\left({S^\prime}^2 + {R^\prime}^2\right)}~,\label{eq1}\\
\frac{\pa S}{\pa t}&=& \frac{{R^\prime}}{\left({S^\prime}^2 + {R^\prime}^2\right)^2}\,\left( {R^\prime}{S^{\prime\prime}}-{S^\prime}{R^{\prime\prime}}\right) + \frac{{R^\prime S^\prime}}{R\left({S^\prime}^2 + {R^{\prime}}^2\right)}~,\label{eq2}
\ea
for the two Cartesian coordinates $R(u,t) = y(u,t)$ and $S(u,t) = x(u,t)$ of the profile curve that generates it.

In contrast to the case of the formation of Type I singularities, and to the best of our knowledge, there exist no mean curvature flow solutions know in closed form that exhibit a Type II singularity. This is also the case for the curve shortening flow of even planar curves for which, apart from the homothetically contracting circle and the  self-similar solutions with intersections of \cite{abresch-langer}, the set of compact known solutions  is exhausted by the oval-shaped solutions that appear in \cite{Angenent:Tori} (paper-clip model \cite{BS, paperclip}). The latter class of solutions can be understood as the mean curvature flow analogue of the Ricci flow Rosenau solution \cite{rosenau:1995} (or sausage model \cite{sausage}). At its final stage,  this compact embedded curve flowing under its mean curvature collapses to a single point while a Type I singularity is formed. Due to the lack of existence of ancient mean curvature flow solutions which can be followed until the occurrence of a Type II singularity, it is necessary to turn to numerical methods  to pursue the investigation of the formation of a Type II singularity during the evolution of a compact surface of revolution. The initial data is provided by the surfaces produced by the revolution of a particular family of closed planar curves. The study can nevertheless be extended by incorporating different families of surfaces with similar asymptotic behavior.

The \textit{Cassini ovals} are planar quartic curves that constitute a generalization of the lemniscate of Bernoulli, and in that respect of the ellipse. A Cassini curve is described as the locus of points
for which the product of the distances to each of two focal points separated by a distance $2a$  equals a constant $b^2$. The corresponding Cartesian equation is
$$
\left(x^2+y^2+a^2\right)^2 - 4 a^2 x^2 -b^4 =0\,.
$$
The shape of the Cassini ovals depends on the ratio $\lambda= \frac{a}{b}$ which will be denoted as shape parameter from now on. Despite their name, these curves do not have an oval shape for generic values of the shape parameter. For $\lambda<1$ the curve is a single loop with an oval  or peanut shape, for $\lambda =1$ the lemniscate curve is produced, while for $\lambda > 1$ they are comprised of two disjointed  loops.
\begin{figure}[h] \centering
\epsfxsize=14cm \epsfbox{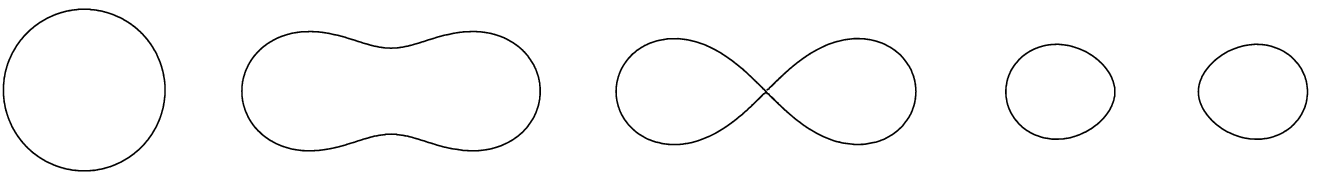}
\put(-388,-6){$\lambda = 0$}
\put(-296,-6){$\lambda < 1$}
\put(-182,-6){$\lambda = 1$}
\put(-64,-6){$\lambda > 1$}
\caption{Cassini curves for different values of the shape parameter $\lambda$.}
\label{s2f1}
\end{figure}

The smooth surfaces of revolution produced by revolving the family of the Cassini curves of a single loop  with $\lambda <1$ around their horizontal axis (Figure \ref{s2f1}),  provide us with the initial data $M_0$. Based on equations (\ref{eq1}) and (\ref{eq2}), the mean curvature evolution $M_t$ is followed without concerning if it is possible to be followed backwards in time to some ancient solution. It should be noted that the resulting rotationally symmetric surfaces are convex for $\lambda \le \frac{1}{\sqrt{2}}$ and, as a consequence, the surfaces $M_t$ evolving under their mean curvature  unavoidably shrink to a point by acquiring a spherical shape, within this range of $\lambda$. On the other hand, the  surfaces generated by Cassini curves  with $\frac{1}{\sqrt{2}} < \lambda <1$, that constitute our variant of a dumbbell surface,  may possibly pinch depending on the precise value of the shape parameter $\lambda$. The detailed numerical evolution for our set-up is presented in section 3, where it is shown that, for a \textit{critical} value of the shape parameter, the flow terminates with the occurrence of a degenerate neckpinch.

\subsection{Effective one-dimensional flow and the translating soliton}

A solution of
the mean curvature flow (\ref{mcf}) that exists for all time and moves by translation in $\R^{n+1}$ is called a translating soliton. More generally, the interest in solitonic solutions of various geometric flows  is, among other reasons,  due to the fact that such solutions model the asymptotic behavior of developing singularities. Regarding the  formation of  Type II singularities, recall the Bryant soliton as a limiting flow of a three-dimensional Ricci flow solution exhibiting such a singularity; or the emergence of the grim-reaper solution during the analogue scenario in the curve shortening flow \cite{angenent:1991, altschuler:1991}\footnote{For a treatment along these lines in relation to the two-dimensional cigar Ricci soliton see \cite{Daska-Sesum}.}. The study of hypersurfaces moving under their mean curvature near a singularity using rescaling  techniques \cite{huisken:Asymptotic} involves  solitonic solutions. The relevant two-dimensional translational solution for the mean curvature flow has been studied among other people by the authors of \cite{grayson, angenent-velaz, Ilmanen, huisken:Asymptotic}, with early numerical evidence for its existence that can be traced in \cite{Stone}. In \cite{altschuler-wu} it was proven that there exists a convex, rotationally symmetric solution translating under its mean curvature with any prescribed constant speed. Unfortunately, an explicit solution describing this soliton in closed form is not available in the literature.

Referring back to the set-up of section 2.1, consider a planar evolving  curve $\Gamma_{t}(x,y)$ that can be represented either as a horizontal graph $y = y(x, t)$, or as a vertical graph $x = x(y, t)$.  As long as the evolution $M_t$ of the  surface of revolution generated by rotating $\Gamma_{0}(x,y)$ around the $x$-axis is regular, the mean curvature flow evolution (\ref{mcfS}) in $\R^3$ is dimensionally reduced to the planar deformation for the graph of the underlying curve $\Gamma_{t}(x,y)$ described
either by the horizontal modified, in reference to the curve shortening flow, graph equation
\be
\frac{\pa y(x,t)}{\pa t} = \frac{y^{\prime\prime}(x,t)}{1+{y^\prime}^2(x,t)} - \frac{1}{y(x,t)}\,,
\label{1dy}
\ee
or equivalently by the corresponding vertical graph equation
\be
\frac{\pa x(y,t)}{\pa t} = \frac{x^{\prime\prime}(y,t)}{1+{x^\prime}^2(y,t)} + \frac{x^\prime(y,t)}{y  }\,,
\label{1dx}
\ee
where prime denotes differentiation with respect to the corresponding spatial variable. The last term of equations (\ref{1dy}) and (\ref{1dx}) captures the $S^1$ extrinsic curvature of the surface due to the revolution around the $x$-axis, and differentiates the resulting  one-dimensional flow from the ordinary curve shortening flow for the graph of its generator.

\begin{figure}[h]
\centering
\vspace{0.25cm}
\begin{minipage}[t]{.35\textwidth}
\begin{center}
\epsfig{file=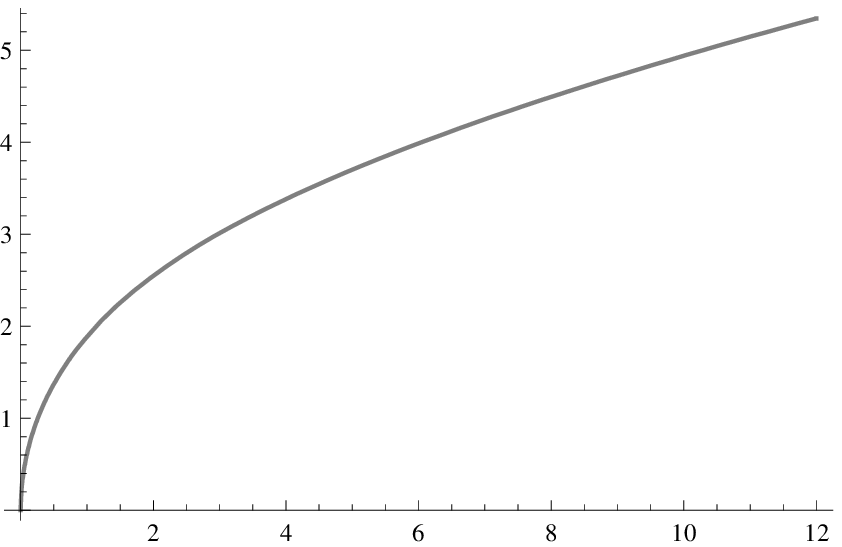, scale=.6}
\put(-80,-36){(\hspace{.1pt}a)}
\put(-150,90){\tiny{y}}
\put(2,-2){\tiny{x}}
\end{center}
\end{minipage}
\hspace{1.cm}
\begin{minipage}[t]{.35\textwidth}
\vspace{-3.32cm}
\begin{center}
\epsfig{file=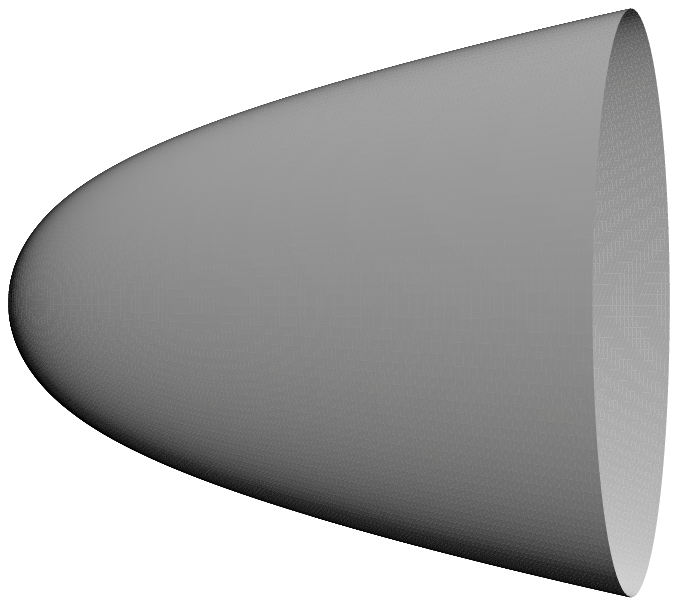, scale=0.8} 
\put(-145,31){(\hspace{.1pt}b)}
\end{center}
\end{minipage}
\vspace{-0.5cm}
\caption{The planar curve (a) that generates the ``bowl" soliton (b) translating with unit speed to the right.}
\label{s2f2}
\end{figure}

Within this effective description of the mean curvature evolution of a rotationally symmetric surface by means of the flow equation of its profile curve (\ref{1dy}) or (\ref{1dx}), a suitable presentation of the corresponding two-dimensional translating solitonic solution follows by considering the possible reparametrizations along this flow. By properly introducing reparametrizations generated by a vector field $\vec{\xi{}}$  to take into account the translational motion of the soliton (see for instance \cite{BS}), the mean curvature flow in $\R^3$ (\ref{mcfS}) acquires the form
\be
\frac{\partial {{\vec{{}X}}}}{\partial t} = -H\,{\hat n} + {\vec{\xi{}}}
\label{mcfSRep}\,. 
\ee
This way, a vector field ${\vec{\xi{}}} = {\vec{\pa}}f(x,y){}$ captures the rigid motion of a \textit{gradient} soliton that translates with a constant velocity $c>0$ along the positive $x$-axis for $f(x)= c x$. Denoting it as a horizontal graph $y = \varphi(x)$, one easily verifies that it satisfies the ordinary differential equation
\be
\varphi(x)\varphi^{\prime\prime}(x) = (1+{{\varphi^{\prime}}^2(x)})\left(1-c\,\varphi(x)\varphi^{\prime}(x) \right)\,,
\label{soliton}
\ee
with respect to the abscissa of the planar curve $x$, or its equivalent one following from (\ref{1dx}) when is preferable to consider a vertical graph. A numerical plot of the solution of (\ref{soliton}) is presented in Figure \ref{s2f2}, along with a $3D$ graphical illustration of the two-dimensional mean curvature flow translating soliton. Alternatively, by canceling the ``freezing" effect of the reparametrization term in (\ref{mcfSRep}), the generator curve of the soliton solution  may be viewed as moving linearly in time along the $x$-direction with constant speed $c$,
\be
y(x,t) =  \varphi(x + ct)\,.
\ee
Likewise, the profile curve that generates a translating
solution along the $y$-axis, arises by reversing the roles of the coordinates $x$ and $y$ and considering vertical graphs instead. Furthermore, the sign of the velocity $c$ selects the orientation of the translational solution in both cases, distinguishing a soliton from an anti-soliton solution. In general, the rotationally symmetric gradient soliton that is generated in this way,  with any line of $\R^3$ as axis of symmetry along which it is translating with a non-zero finite velocity $c$, has a ``tip" at its intercept with that line and asymptotically opens to a hyperboloid.

\begin{figure}[h] \centering
\epsfxsize=7cm \epsfbox{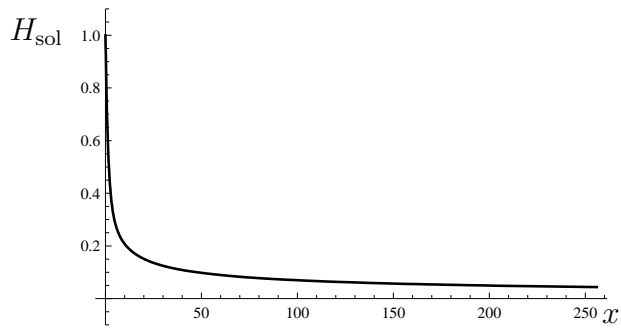}
\put(-226,108){$\small{H_{\rm{sol}}}$}
\put(-2,1){$\small{x}$}
\caption{Mean curvature of the unit speed translating soliton in terms of the horizontal distance from its tip.}
\label{s2f3}
\end{figure}

A convenient way to describe the mean curvature of the soliton under investigation is to introduce the
slope of the curve that produces it via rotation.
In general, the mean curvature $H$ of a smooth rotationally symmetric surface generated by
revolving a profile planar curve $y(x)$ around the $x$-axis is related to the mean curvature of its generator curve $H_{c}$ as
\be
H = H_c + \frac{{\rm cos}\beta}{y}\,,
\ee
where $\beta$ is the angle formed by the tangent at each point of the curve and the $x$-axis, i.e.,
\be
\beta = {\rm arctan}\,y^{\prime} (x)~,
\label{slope}
\ee
when viewed as a horizontal graph. In terms of this angle, the mean curvature of the soliton translating along the positive $x$-axis with velocity $c$
acquires the simple form
\be
H_{\rm{sol}} = c\, {\rm sin}\beta \qquad \mbox{with} \qquad 0 \le \beta \le \pi\,,
\label{Hsol}
\ee
where the defining equation (\ref{soliton}) was used. It is instructive to realize that  the result (\ref{Hsol}) coincides with the mean curvature of the translating soliton of the curve shortening flow on the two-dimensional Euclidean plane, the grim-reaper solution \cite{BS}. Put differently, the planar curve that satisfies (\ref{soliton}) is such that the rotation around the $x$-axis produces just the curvature required so that the  generated non-compact soliton surface has mean curvature, in terms of its slope,  matching the one of its one-dimensional cousin. The expression  (\ref{Hsol}) for the mean curvature of the two-dimensional soliton, given here in closed form in terms of the slope with respect to its axis of symmetry, corresponds to a static solution of equation (\ref{mcfH}) when reparametrizations are taken into account. Equation (\ref{mcfH}) may in principle be used to study the stability properties of the rotationally invariant translating soliton solution \cite{stability},  along the lines of \cite{BS}. For completeness, in Figure \ref{s2f3},  the mean curvature of the soliton is presented as a function of the horizontal distance from its tip.
The inspection of this plot reveals that  for the soliton (\ref{soliton}), the mean curvature is everywhere positive, acquires its maximum value at the tip ($\beta = \pi/2$) and monotonically dies off to zero ($\beta \to 0$). This is in full accordance with a general
theorem stating that any eternal strictly convex solution of the mean curvature flow with the mean curvature attaining a maximum value at only one point,
must necessarily be a translating soliton \cite{hamilton:1995}. This is an important ingredient that goes into the study of Type II singularities by means of rescaling techniques with the soliton solutions coming into play as  limiting flows, which will be discussed below. In passing, we note that modulo direction, orientation and translational speed, the soliton depicted in Figure \ref{s2f2} is the unique two-dimensional translating soliton solution that can be generated by rotating a horizontal with respect to the axis of revolution graph. It must be stressed though, that other two-dimensional mean curvature flow translating solitons, not falling into this category, do exist\footnote{We thank G. Huisken for pointing out their existence and also clarifying their possible relation to the formation of Type II singularities in the course of the mean curvature flow evolution of \textit{immersed} (hyper)surfaces (see also \cite{Wang}).}.

\subsection{Rescaling the singularity}

The asymptotic description of a mean curvature flow solution near a singularity of both types, through the examination of sequences of rescaled solutions, involves the notion of the so-called \textit{blow-ups}. Consider the evolution $M_t$ of a compact, smooth $n$-dimensional hypersurface  of positive mean curvature for $t \in [0, T)$, where $T$ is the first singular time. Let $p_0$  be the point of $M_T$  at which the curvature becomes unbounded as $t \to T$ (the blow-up point), and the forming singularity be of Type II. If for every integer $k\ge1$, $t_k \in [0,T - \frac{1}{k}]$ and $p_k \in M_t$,  $\{p_k,t_k\}$ is a sequence such that
\be
H^2(p_k,t_k)(T-\frac{1}{k}-t_k) =  \displaystyle \max_{t\le T -1/k,\, p\,\in M_t}H^2(p,t)(T-\frac{1}{k}-t)
\ee
then the sequence of the rescaled flows of $\vec{{{}X}}: M_t \to \R^{n+1}$,
\be
\vec{{{}X}}_{(k)}(p,\tau) = \e_k \,\left(\vec{{{}X}}(p,t_k+\e^{-2}_k\tau) - \vec{{{}X}}(p_k,t_k)\right) \quad \mbox{for} \quad \tau \in [- \e_k^2\,t_k,\e^2_k(T-t_k-\frac{1}{k})]\,,
\label{bupII}
\ee
where
\be
\epsilon_k = H(p_k,t_k) \,,
\ee
converges smoothly to a mean curvature eternal flow. The limiting hypersurface is weakly convex and has uniformly bounded mean curvature \cite{huisken-sine99}. Section 3 provides numerical evidence that the type II blow-up (\ref{bupII}) of the mean curvature flow of the dumbbell-   shaped surface of revolution generated by the Cassini curve with shape parameter equal to the critical value $\lambda_{c}$, with either of its two poles as blow-up point, converges to the rotationally symmetric translating soliton (\ref{soliton}).

In the same section, there is also a focus on the limiting flow of the central region of the neck  of the evolving surfaces where the pinching is observed for generic supercritical values of $\lambda$. In \cite{huisken:Asymptotic} (but see also \cite{souganidis}) it was shown that the limiting flow of the parabolic, or type I, blow-up of a rotationally symmetric shrinking neck converges to the cylinder of radius $\sqrt{2}$. In terms of the profile curve that generates such a surface by rotation, this amounts to
\be
\frac{1}{\sqrt{T-t}} \,y\left(\frac{x}{\sqrt{T-t}}\right) \rightarrow \sqrt{2} \qquad \mbox{as} \quad t \to T\,.
\label{bupN}
\ee
The continuous rescaling (\ref{bupN}) is relevant for the asymptotic description of the
generic neckpinch that occurs when the regular evolution of our supercritical surfaces comes to an end with the formation of a Type I singularity. For the case of a degenerate neckpinching a more refined rescaling is needed. This is provided in \cite{angenent-velaz}, and in terms of the profile curve it has the form
\be
\tilde{y}(\tilde{x}) = \sqrt{2} + K\, {\rm{Hm}}_m(x) (T-t)^{\frac{m}{2}-1} + {{\mathcal{O}}}\left((T-t)^{\frac{m}{2}-1}\right)\quad \mbox{as} \quad t \to T\,,
\label{angenent-velazDeg}
\ee
with
\be
\tilde{y} = \left(T - t\right)^{-1/2} y\, \quad \mbox{and} \quad \tilde{x} = \left(T - t\right)^{1/2} x\,,
\ee
where ${\rm{Hm}}_m(x)$ is $m^{\rm{th}}$ Hermite polynomial normalized such that the $x^m$ term has unit coefficient. For a left-right symmetric surfaces of revolution, the integer $m$ appearing in (\ref{angenent-velazDeg}) is even and greater than two. Also, in that case,  the numerical coefficient $K$ is determined based on the distance between the Type II singular point and the poles of the surface.

\section{Numerical investigation}

In this section the results of the numerical simulations of the mean curvature flow of the chosen   one-parameter family of rotationally symmetric dumbbell surfaces in $\R^3$ are presented. The initial configurations are provided by the compact surfaces of revolution generated by revolving the one-component Cassini curves around their horizontal axis of symmetry, as described in section 2.  These are parametrized in terms of the shape parameter, with $0 \le \lambda < 1$. For small values of this parameter (``loose corsetting" relative to a round sphere in the terminology of \cite{G-I 2003, G-I 2007}), even if the initial data is not convex, the surfaces evolving under their mean curvature  become convex and asymptotically approach the geometry of the round two-sphere. In contrast, the dumbbell surfaces with large values of the shape parameter $\lambda$ (``tight corsetting") do not loose their neck until their evolution becomes singular with the development of a generic neckpinch at the center of their left-right symmetric extend.
\begin{figure}[h] \centering
\epsfxsize=8cm \epsfbox{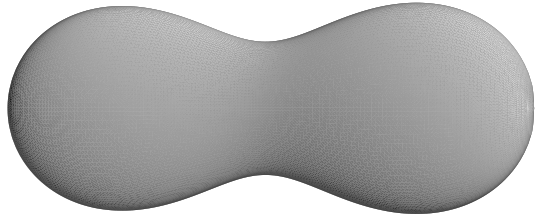}
\caption{A generic dumbbell-shaped supercritical initial surface.}
\label{s3f1}
\end{figure}
The numerical simulation of the mean curvature flow of such initial data with two kinds of possible asymptotic behavior depending on the shape parameter $\lambda$, reveals the existence of a critical  transition value $\lambda_{c}\approx 0.9076$ for which a degenerate neckpinch occurs. Unlike both subcritical ($\lambda < \lambda_{c}$) and supercritical ($\lambda > \lambda_{c}$) initial surfaces, the smooth evolution of the dumbbell with $\lambda = \lambda_c$ terminates with the formation of a Type II singularity. Focusing on the flow of this critical initial data, it is possible to verify that  the critical surface in the course of its evolution becomes increasingly cylindrical in the neighborhood of the initial pinching and that the mean curvature blows up at its two poles. This is in full analogy to the critical flow of \cite{G-I 2003, G-I 2007} for the Ricci flow. It is valuable to stress that, at its very final stage, the critical dumbbell surface collapses to a point but without turning convex until that time. The close examination of the limiting flow, with either of the two poles as a blow-up point, provides strong numerical evidence that it is modeled with high accuracy by the two-dimensional translational mean curvature soliton (\ref{soliton}).

The mean curvature evolution of this one-parameter family of surfaces is studied by numerical means. One of the parameters of the Cassini closed curves that generate them is held fixed, $b=1$, and the other one takes values in the interval $[0,1)$.  This way, the induced metric of the initial data is parametrized in terms of the scale parameter  $\lambda$,
\be
ds^2 = T_0^2(\theta\,;\lambda)\,d \theta^2 + R_0^2(\theta\,;\lambda)\,d \phi^2 ~,
\ee
where $T_0^2(\theta\,;\lambda) = {S_0^\prime}^2(\theta\,;\lambda) + {R_0^\prime}^2(\theta\,;\lambda)$ with  $0\le \lambda < 1$ and $0 \le \theta  \le \pi $. The metric components are numerically evolved based on the mean curvature evolution equations (\ref{eq1}) and (\ref{eq2}) for the two planar coordinates $S(\theta,t)$ and $R(\theta,t)$ of the profile curve as functions of the angle $\theta$ and the evolution time $t$. Angular derivatives are approximated by centered finite differences with an angular step size $\Delta\theta = \pi/N$, where $N$ is the number of  angular grid points $\theta_i$ employed. As in \cite{G-I 2003, G-I 2007} the angular grid is augmented by considering two external fictitious points $\theta_0$  and $\theta_{N+1}$ on both of its ends. Periodic boundary conditions are implemented by imposing  the conditions $S_0(t=0) = S_1(0)$, $S_{N+1}(0) = S_N(0)$ and $R_0(0) = - R_1(0)$, $R_{N+1}(0) = - R_N(0)$ at the two poles. In order to meet the need of increased accuracy close to the poles of the evolving surface ($\theta = 0$ and $\theta = \pi$), especially when approaching the forming singularity in time, variable temporal step size is used  when the differentiation with respect to time in (\ref{eq1}) and (\ref{eq2}) is approximated by finite differences. Computer code was executed in \textit{Mathematica} with up to 10,000 angular grid points. This was reduced by a factor of two when the shape parameter of the initial surface was adjusted close to its critical value in order to avoid numerical instabilities due to the stiffness of the simulated system of finite difference equations. In general,
the results of repeated runs of the computer code clearly differentiate the asymptotic behavior of  subcritical and supercritical initial geometries in a well distinguishable way.

\subsection{Critical flow and  comparison with the soliton}

As in \cite{G-I 2007},  two complementary comparison methods are presented in order to confirm the claim that
the flow of the critical initial surface is asymptotically modeled by the mean curvature translating soliton.

\begin{figure}[h] \centering
\epsfxsize=15cm \epsfbox{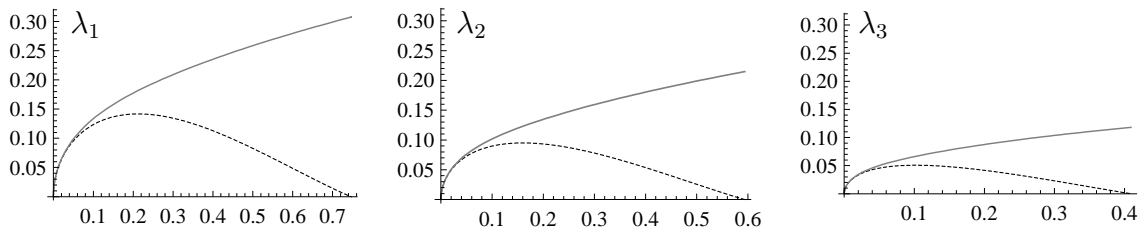}
\put(-404,76){$\lambda_1$}
\put(-258,76){$\lambda_2$}
\put(-106,76){$\lambda_3$}
\caption{The profile curve for three supercritical surfaces at pinching (dashed lines) compared to the soliton curve (solid lines). Only the part from the left pole up to the center is depicted and $\lambda_1 =0.94,\lambda_2 = 0.93, \lambda_3 =0.92 $.}
\label{s3f2}
\end{figure}

As a first examination approach,  the geometry of an appropriate mean curvature soliton solution is compared with the geometries near the left pole for a sequence of supercritical initial
geometries, with shape parameters approaching the critical value $\lambda_c$, at the time of maximum curvature. The soliton's speed parameter $c$ (cf. section 2.2) is determined by requiring in each case the mean curvature at its tip to match the maximum value of the mean curvature  at the poles of the evolving dumbbell surfaces. Note that during the mean curvature evolution of these surfaces, the maximum value of the curvature at the poles is attained when a generic neckpinch occurs at their center rendering any evolution thereafter singular. A comparison is presented in Figure \ref{s3f2}, where the profile curve that generates the corresponding soliton (solid line) is plotted against the profile curve that, by revolution, produces the final phase of three different supercritical initial geometries while their neck is pinching (dashed lines) versus the horizontal distance from their left pole up to their center. Due to the unequal spatial extend of the dumbbells at this singular time, three separate graphs are depicted. Note that the translational speed parameter of the soliton increases from left to right.
By inspection of this set of graphs, it is clear that as the shape parameter $\lambda$ of the initial data decreases towards the critical value, the surface near the pole of the final stage of the singular dumbbell  tends to approach the surface of the corresponding mean curvature soliton solution. Recall that, in this comparison test, no magnification  of the dumbbell's surface has been performed during its evolution. The accuracy of the polar modeling of the dumbbell by the soliton increases for initial data corresponding to slightly supercritical shape parameter values. Evidently, owing to the reflection symmetry of the family of the initial configurations, a similar behavior is observed when the flow close to the right pole is examined.

\begin{figure}[h] \centering
\epsfxsize=8cm \epsfbox{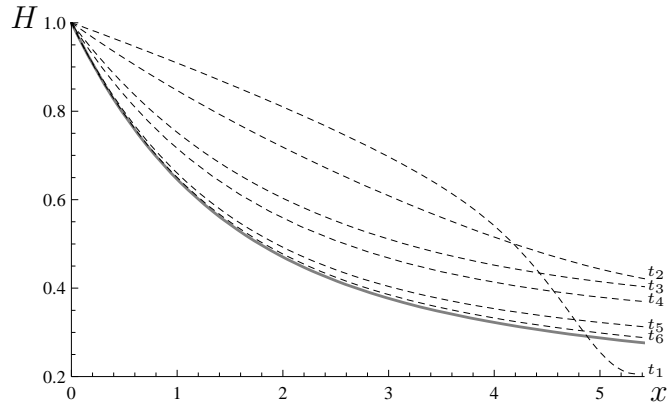}
\put(1,11){\tiny $t_1$}
\put(1,48){\tiny $t_2$}
\put(1,43){\tiny $t_3$}
\put(1,38){\tiny $t_4$}
\put(1,28){\tiny $t_5$}
\put(1,24){\tiny $t_6$}
\put(-240,142){$\small{H}$}
\put(2,1){$\small{x}$}
\caption{Mean curvature near the pole of the rescaled critical surface for six subsequent times $t_i$ compared to the mean curvature of the soliton (single solid line).}
\label{s3f4}
\end{figure}

In order to obtain an accurate description of the critical flow corresponding to an initial data with a critical shape parameter value $\lambda_c$, a second comparison analysis is performed. This time, a choice for a critical initial surface is made with shape parameter as close to its critical value as permitted by the numerical determination of $\lambda_c$.
The flow of the dumbbell with $\lambda \simeq \lambda_c$ is simulated and it is observed that the maximum curvature at the poles increases along the flow of this close to critical initial surface. It eventually diverges when $\l$ is fine-tuned, as attainable, to be critical. Next, a sequence of times approaching the time of singularity formation is considered. For each chosen instant of evolution time, the blow-up geometry as specified in section 2.3 is determined, with one of the two poles as blow-up point. This way the rescaled  dumbbell surface  has maximal curvature at the left, for example, pole  equal to one. This is repeated for the remaining  subsequent time  instances of the critical flow. The results of this procedure are graphically  demonstrated  in the following figures. In Figure \ref{s3f3}, the blow-up of the profile curve for six different times of the evolution of the ``critical" initial surface
are plotted in comparison with the profile curve that generates the mean curvature steady soliton (solid line) corresponding to a translational speed of unit value to the right, in terms of the horizontal distance from the tip of the soliton.
\begin{figure}[h] \centering
\epsfxsize=7.9cm \epsfbox{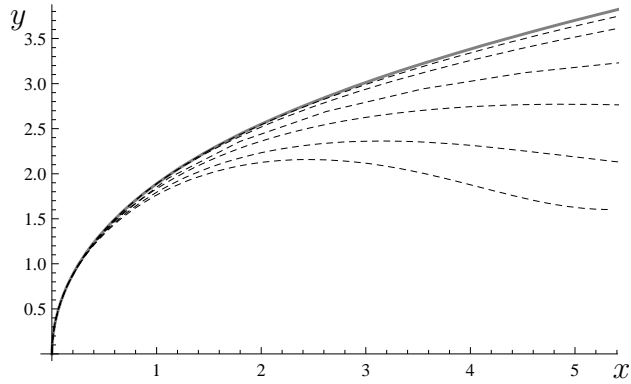}
\put(-230,135){$\small{y}$}
\put(-2,1){$\small{x}$}
\caption{ A sequence of subsequent rescaled profile (dashed) curves along the critical flow with time increasing upwards compared to the unit speed soliton profile (solid) curve.}
\label{s3f3}
\end{figure}
It is observed that the rescaled curves near the pole approach the profile curve of the unit speed soliton solution with increasing agreement along the flow of the critical surface. The presentation of the results of this simulation is supplemented by the graphical comparison of the mean curvature near the poles for a sequence of  rescalings of the critical flow to the mean curvature close to the tip of the steady soliton of Figure \ref{s3f4}. The simulated mean curvature for subsequent instances of time  is plotted as a function of the radial distance and excellent agreement with the mean curvature of the soliton is detected at later times.
\vspace{4pt}

The results of both of the above described comparison tests provide convincing evidence that the limiting flow of the critical initial geometry lying at the ``transition point'' of our one-parameter family of initial surfaces, with different mean curvature asymptotic behavior on both sides,
is accurately modeled by the translating soliton of section 2.

\subsection{Degenerate neckpinching}

The accurate modeling of the polar flow by the surface of the appropriate gradient soliton strongly supports the identification of the singularity along the critical flow as a Type II. A supplementary verification  may also be produced through the detailed study of the limiting flow, not close to the poles, but close to the center of the evolving critical surface instead. The occurrence of a mean curvature flow degenerate neckpinch can indeed be confirmed through its asymptotic description.

\begin{figure}[h] \centering
\epsfxsize=15.75cm \epsfbox{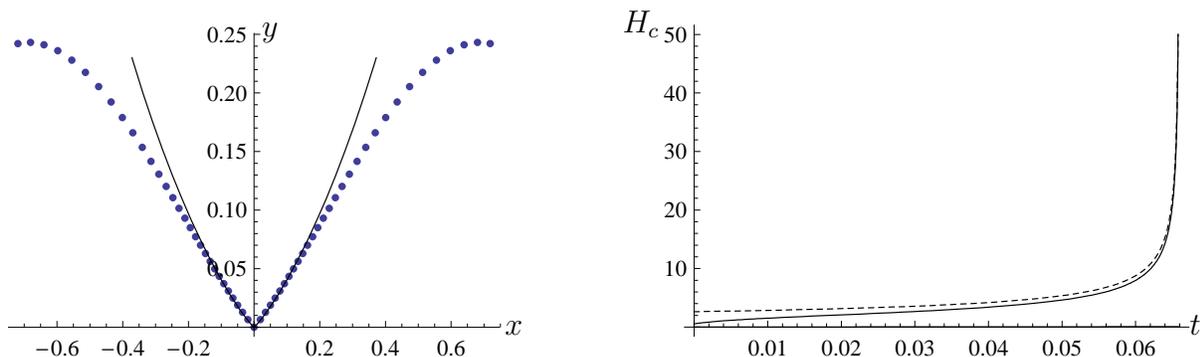}
\put(-352,124){$y$}
\put(-260,11){$x$}
\put(-215.5,125.5){$H_{c}$}
\put(-1,11){$t$}
\caption{ The cusp singularity of a generic neckpinch for $\lambda=0.96$; one every 100 grid points is depicted out of 10,000 used in total for the simulation. On the right the mean curvature of the center point (solid line) is compared to the asymptotic expression (\ref{angenent-velazH}).}
\label{s3f5}
\end{figure}

Certainly, the decisive factor of the classification of a mean curvature singularity is the blow-up rate of the maximal mean curvature. Based on the results of \cite{angenent-velaz}, the mean curvature of the singular point of a rotationally symmetric surface has a limiting  form
\be
H = \left(\frac{\xi}{m} + {{\mathcal{O}}}(1) \right)(T-t)^{1/m-1}\qquad \mbox{as} \quad t \to T\,,
\label{angenent-velazH}
\ee
where for the case of a left-right symmetric surface $m$ is a positive \textit{even} integer and $\xi$ a constant determined by the distance of the poles of the surface from its center at any instant of time close to the singular time $T$. In the case of a Type II singularity, the divergence of the mean curvature at the pole of a surface of revolution with one neck is expected to be reproduced by the expression (\ref{angenent-velazH}) for $m=4$.  On the other hand, a Type I singularity corresponds to $m=2$ in agreement with the rate (\ref{Arate}). Referring to the numerical simulations, recall that the evolution of all supercritical initial data becomes singular with the occurrence of a generic neckpinch at their center ($x=0$). According to \cite{angenent-velaz} (section 9), the asymptotic shape of the curve that produces  this generic neckpinch by rotation, while a Type I singularity is formed, is described  by
\be
y(x,T) = K \frac{|x|}{\sqrt{\log\frac{1}{|x|}}}\,,
\label{genericP}
\ee
where again the constant $K$ is determined by the distance of the poles from the center and $T$ is the singular time. In Figure \ref{s3f5}, a graphical comparison of the limiting shape of the central region of the profile curve for a typical supercritical surface and the generic pinching (\ref{genericP}) is presented.  Adequate agreement is observed as $x \to 0$.
Moreover, the Type I characterization of this singularity is also confirmed by the second graph of the same figure, where the mean curvature of the central point is compared to the limiting blow-up rate (\ref{angenent-velazH}) with $m=2$. In both graphs, the comparison is extended beyond the expected agreement.

A direct numerical verification of the formation of a Type II singularity  at the final stage of the critical surface based on the expected blow-up rate (\ref{angenent-velazH}) is essentially impossible for two reasons. First of all, for the one-parameter family of initial geometries selected here, a degenerate neckpinching occurs for a single value of the shape parameter $\lambda$. As a consequence, the numerical distinction between different blow-up rates (\ref{angenent-velazH}) with $m=2$ and $m=4$ requires an input of very high precision for the critical value $\lambda_{c}$. Second, numerical implications due to the left-right symmetry of the surfaces used in the present set-up also prevent such a resolution.  Specifically, the coefficient $\xi$ in (\ref{angenent-velazH}) acquires a vanishing limiting value, since the critical surface collapses to a point as the singularity of the second type is formed.

However, numerical evidence in favor of the occurrence of a degenerate neckpinch along the critical flow  may be produced by examining the asymptotic shape of the central region of the critical surface. This is achieved by considering a sequence of subcritical initial surfaces with shape parameter close to the critical value and focusing on their central region. During their mean curvature evolution these surfaces remain non-convex until a time just before their collapse. The fine examination of the asymptotic central shape of these slightly subcritical surfaces just before they turn convex and shrink to a point soon after, reveals a match to the degenerate neckpinching of equation (\ref{angenent-velazDeg}).
\begin{figure}[h] \centering
\epsfxsize=7cm \epsfbox{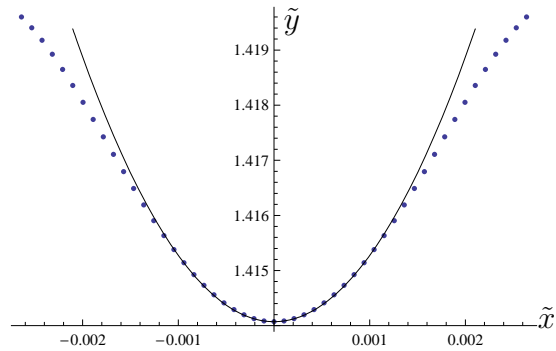}
\put(-96,121){$\tilde{y}$}
\put(+0.2,8){$\tilde{x}$}
\caption{ A simulation of the degenerate neckpinch for $\lambda = 0.9076$ with one every 50 grid  points depicted out of a total 5,000 in comparison with the blow-up of the curve (\ref{angenent-velazDeg}).}
\label{s3f6}
\end{figure}
In Figure \ref{s3f6}, the central part of the curve that generates such a surface of revolution close to its extinction time $T$ is compared to the blow-up of the degenerate neckpinch (solid line) described by equation   (\ref{angenent-velazDeg}). This plot depicts the generator curve at $t = 0.99 T$ (i.e., a vertical magnification factor of the order $\approx 50$) with initial shape parameter $\lambda = 0.9076$. In this case, improved agreement is detected as the critical shape parameter is approached from below, to the extend permitted by the numerically determined value of  $\lambda_{c}$,  and at later time. This supports the claim that the critical initial surface evolves at its final stage into a degenerate neckpinch. Actually, this reasoning can be inverted and used to improve the numerical value of $\lambda_{c}$ by demanding accurate modeling of the limiting form of the central region by equation (\ref{angenent-velazDeg}) for the critical surface at the extinction time. A precise determination of the critical value of the shape parameter based on non-numerical alternative methods would in general be very helpful.

\begin{figure}[h] \centering
\epsfxsize=9cm \epsfbox{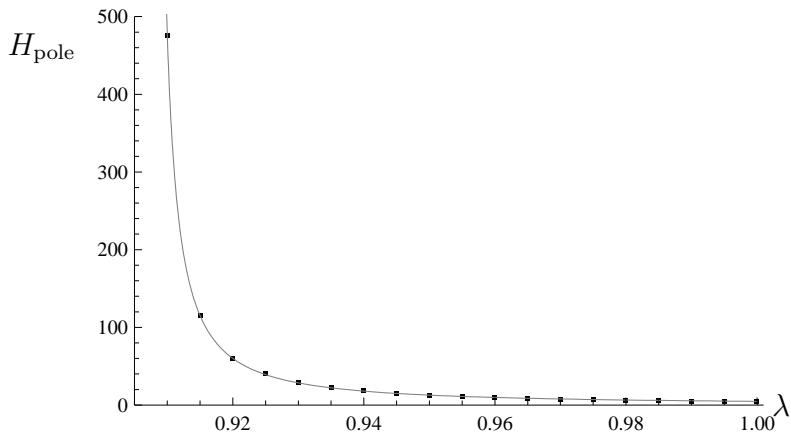}
\put(-290,144){$H_{{\rm pole}}$}
\put(-1,7){$\lambda$}
\caption{Maximal mean curvature of the poles as a function of the shape parameter.}
\label{s3f7}
\end{figure}

This presentation of the numerical results concludes with a graphical representation of the dependence of the mean curvature maximum value at the poles of the family of initial surfaces on the shape parameter $\lambda$.  For all supercritical surfaces, the mean curvature at the poles attains its maximum value at the occurrence of a generic pinching at their center. These values are depicted in Figure \ref{s3f7} for the supercritical range of the shape parameter. It is observed that the mean curvature maximum value increases as the shape parameter decreases. The depicted divergence as the critical value is approached from above is attributed to the Type II singularity of the degenerate neckpinch. The numerically extracted values are reproduced by an expression of the form
\be
H_{\rm{pole}} =  \Lambda_0 \left(\lambda - \lambda_{c} \right)^{-n}\,,
\label{criticalH}
\ee
where $\Lambda_0$ is a constant that depends on the choice of the one-parameter family of initial surfaces. For the surfaces of revolution generated by the Cassini curves, the value of the mean curvature at the poles of the surface of revolution generated by the lemniscate curve ($\lambda =1$) may be used as an input. The simulation of the mean curvature flow of this family of initial surfaces parameterized by $\lambda$  suggest a value $n\approx 1.26$ for the ``critical exponent" of (\ref{criticalH}). It must be stressed though, that this numerical value highly correlates with the value used for $\lambda_{c}$.
%

\section{Conclusions and outlook}

In this work we have performed a numerical study of the mean curvature evolution of two-dimensional surfaces of revolution of a dumbbell shape\footnote{For an early computer study of a collapsing dumbbell under mean curvature flow, that motivated investigations of developing singularities before a generic embedded surface shrinks to a point \cite{grayson}, see \cite{Sethian}.}. We considered a specific one-parameter family of initial reflection symmetric surfaces generated by revolving the Cassini curves around their axis of symmetry. By  simulating the mean curvature flow of this family of initial data, parametrized in terms of the shape parameter $\lambda$ of the Cassini curves, we succeeded in revealing a behavior in the emerging singularity pattern that resembles the critical behavior of gravitational collapse. In particular, all initial surfaces with a shape parameter less than a critical value $\l_c$ collapse to a point in finite time by acquiring spherical shape. On the other hand, the evolution of the initial dumbbell surfaces with supercritical shape parameter becomes singular with the formation of a generic neckpinch at their center. The fine examination of the flow of the critical surface allows for the detection of the occurrence of a Type II singularity. Compared to the approach of \cite{G-I 2003, G-I 2007} on Ricci flow critical behavior, we note that in the present work  a normalized variant of the mean curvature flow was not employed. The area preserving flow of \cite{huisken:Asymptotic}, or the modified flow \cite{NVflow} that preserves the enclosed volume, could have been incorporated into our study without altering its results. The use of the ordinary (unnormalized) mean curvature flow permitted a direct comparison with the results of \cite{angenent-velaz} on Type II mean curvature singularities and allowed for a numerical verification of the formation of a degenerate neckpinch at the final stage of the flow of the critical surface.

The numerical results of the mean curvature simulation of the rotationally symmetric family of our initial surfaces provide strong evidence that the polar critical flow is accurately modeled by the rotationally invariant translating soliton solution (\ref{soliton}). We are able to extend this study
by incorporating the simulation of different families of initial compact rotationally symmetric surfaces, including the one used in \cite{G-I 2003} and \cite{G-I 2007}.  In all   cases examined, mean curvature flow critical behavior is in general observed together with the occurrence of a degenerate neckpinch. Critical behavior is also detected by direct numerical simulations of initial configurations which are not left-right symmetric. In this case, the part of the surface on one side of the degenerately collapsing neck shrinks to zero size while a Type II singularity is formed. In addition, a similar behavior is observed for initial data with more than one neck without any clear evidence of coalescence. Further studies are nevertheless required before any universality claims \`{a} la Choptuik can be made about critical behavior at the threshold of mean curvature singularity formation.

A  relevant future direction of investigation would be to consider the evolution of non-rotationally symmetric surfaces within the framework of the current study. This is a very challenging and difficult task. As an alternative direction, we suggest the investigation of whether critical behavior is present in the mean curvature evolution of torus-like initial configurations, as opposed to the  dumbbell-like examined in this work (\cite{souganidis} may turn out to be helpful in this respect). If critical behavior is detected for this alternative class of initial mean curvature data, the comparison of the corresponding critical limiting flow with the one of the degenerately collapsing critical dumbbell of the present work may reveal the mean curvature analogue of the scale coordinate of gravitational critical phenomena \cite{Choptuik, Gundlach}.

Mean curvature flow is the basic example of an extrinsic curvature flow. Critical behavior and similarities with Choptuik scaling can be sought for other extrinsic flows. Taking into account the important role of the inverse mean curvature flow in the proof of the Riemannian analogue
of the Penrose inequality of general relativity \cite{inverseMCF}, we believe the investigation of critical behavior of this flow deserves an independent study. Such a study can motivate the introduction of the appropriate functionals with power-law scaling even for the mean curvature flow and may also provide a direct link between critical behavior at the threshold of singularity formation of curvature flows and critical behavior at the threshold of black hole creation.

An essential ingredient of the presentation of our study was the translating mean curvature gradient soliton that  deserves further consideration. Anticipating the role of the grim-reaper (hair-pin) solution, as the curve shortening flow translating soliton, in the construction of integrable quantum field theories of boundary interactions \cite{paperclip}, and also, the role of translating solitons in the level-set approach of the mean curvature flow \cite{Ilmanen} we indicate
that work is in progress towards finding the proper placement of this solitonic surface within the context of boundary conformal field theory. Finally, also interesting is  the investigation of generalizations of this solitonic solution for non-flat ambient spaces \cite{BS} and/or in the presence of anti-symmetric tensor fields, motivated by string theory \cite{Bakas, Wool}, that may influence the structure of singularity formation.
\vskip1.5cm
\centerline{\bf Acknowledgements}
{}\hskip -0.25in The research of C.S. is partially supported by a I.K.Y. funding for postdoctoral research by the Greek government. Numerous discussions with Ioannis Bakas on geometric flows have been very helpful and inspiring. We want to thank the organizers of ``Field Theory and Geometric Flows" workshop in Munich, November 2008, for providing a stimulating atmosphere that motivated this work, and J. Isenberg for explaining aspects of his work.

\end{document}